\def\sqr{\sqrt{1-\ee}}
\def\wq{\ton{\kx^2 - \ky^2}}

\def\cp{\ton{3 + \cII}}
\def\wl{\ton{\kx^2 + \ky^2 - 2 \kz^2}}
\def\wpl{\ton{\kx\cO + \ky\sO}}
\def\wmn{\ton{\ky\cO - \kx\sO}}
\def\ple{\ton{- 2 + \ee + 2 \sqr}}

\def\nk{n_{\rm b}}

\def\rfr#1{eq. (\ref{#1})}

\def\derp#1#2{\rp{\partial{#1}}{\partial{#2}}}

\def\eqi{\begin{equation}}
\def\eqf{\end{equation}}
\def\eqia{\begin{eqnarray}}
\def\eqfa{\end{eqnarray}}
\def\Om{\mathit{\Omega}}
\def\rp#1#2{{#1\over#2}} \def\lb#1{\label{#1}}

\def\kx{\hat{k}_x}
\def\ky{\hat{k}_y}
\def\kz{\hat{k}_z}
\def\bds#1{\boldsymbol{#1}}

\def\coo{\cos 2\omega}
\def\soo{\sin 2\omega}
\def\cO{\cos\Om}
\def\su{\sin u}
\def\cu{\cos u}
\def\sO{\sin\Om}
\def\cOO{\cos 2\Om}
\def\sOO{\sin 2\Om}

\def\cI{\cos I}
\def\sI{\sin I}
\def\cII{\cos 2I}
\def\sII{\sin 2I}

\def\cu{\cos u}
\def\su{\sin u}

\def\ee{e^2}

\def\ton#1{\left(#1\right)}
\def\qua#1{\left[#1\right]}
\def\grf#1{\left\{#1\right\}}
\def\ang#1{\left\langle #1\right\rangle}

\documentclass[useAMS,usenatbib]{mn2e}
\usepackage{amsmath}
\usepackage{amscd}
\usepackage{amssymb}
\usepackage{txfonts}
\RequirePackage{color}

\allowdisplaybreaks[1]

\title[Local Position Invariance and planetary precessions]{\textcolor{black}{Preliminary bounds} of the gravitational Local Position Invariance from Solar System planetary precessions}

\author[L. Iorio]{L.
Iorio$^{1}$\thanks{E-mail:
lorenzo.iorio@libero.it}\\
$^{1}$I Ministero dell'Istruzione, dell'Universit\`{a} e della
Ricerca (M.I.U.R.), Viale Unit\`{a} di Italia 68
Bari, (BA) 70125,
Italy}

\begin{document}

\maketitle

\label{firstpage}

\begin{abstract}
In the framework of the Parameterized Post-Newtonian (PPN) formalism, we calculate the long-term Preferred Location (PL) effects, proportional to the Whitehead parameter $\xi$, affecting  all the Keplerian orbital elements of a localized two-body system, apart from the semimajor axis $a$. They violate the gravitational Local Position Invariance (LPI), fulfilled by General Relativity (GR). We \textcolor{black}{obtain preliminary bounds on} $\xi$ by using the latest results in the field of the Solar System planetary ephemerides. The non-detection of any anomalous perihelion precession for Mars allows us to indirectly infer $|\xi|\leq 5.8\times 10^{-6}$. \textcolor{black}{Such a bound is close to the constraint, of the order of $10^{-6}$, expected from the future BepiColombo mission to Mercury. As a complementary approach, the PL effects should be explicitly included in the dynamical models fitted to  planetary data sets to estimate $\xi$ in a least-square fashion in a dedicated ephemerides orbit solution.} The ratio of the anomalous perihelion precessions for Venus and Jupiter, determined with the EPM2011 ephemerides at the $<3\sigma$ level, if confirmed as genuine physical effects needing explanation by future studies, rules out the hypothesis $\xi \neq 0$. A critical discussion of the $|\xi| \lesssim 10^{-6}-10^{-7}$ upper bounds obtained in the literature from the close alignment of the Sun's spin axis and the total angular momentum of the Solar System is presented.
\end{abstract}

%
%


 \maketitle

\section{Introduction}
In the framework of the Parameterized Post-Newtonian (PPN) formalism \citep{1968PhRv..169.1017N,1971ApJ...163..611W,1972ApJ...177..757W,1993tegp.book.....W}, the Whitehead parameter $\xi$ accounts for a possible Galaxy-induced anisotropy affecting the dynamics of a gravitationally bound two-body system in the weak-field and slow-motion approximation \citep{1973ApJ...185...31W}. Indeed, non-zero values of the parameter $\xi$  determine the level to which  Preferred Location (PL) effects occur in an alternative gravitational theory \citep{1993tegp.book.....W}; in General Relativity (GR), which fulfils the Local Position Invariance (LPI), it is $\xi_{\rm GR} = 0$.
To date,  $\xi$ has been constrained, in the weak-field regime, with superconducting gravimeters on the Earth's surface to the $\sim 10^{-3}$ level \citep{1976ApJ...208..881W}, and with the Sun's spin axis to the $\sim 10^{-7}$ level \citep{1987ApJ...320..871N}. Although not explicitly taken into account so far in the Lunar Laser Ranging (LLR) analyses \citep{LLR08},   $\xi$ may be constrained with such a technique down to the $\sim 10^{-5}$ level \citep{2013arXiv1307.2637S}. The pulse profiles of some solitary radio pulsars have recently allowed to constrain the strong-field version $\hat{\xi}$ of the Whitehead parameter down to the $\sim 3.9\times 10^{-9}$ level \citep{2013arXiv1307.2637S}. A less stringent constraint, i.e. $|\hat{\xi}|< 3.1\times 10^{-4}$, was obtained from the spin evolution of two binary pulsars using probabilistic arguments \citep{2012arXiv1211.6558S}.
An overview on the current constraints on all the PPN parameters can be found in \citet{2001LRR.....4....4W,2010IAUS..261..198W}.

In this paper, we use the latest advances in the field of the Solar System planetary ephemerides to put new constraints on $\xi$ and to critically discuss some of the previous ones. The plan of the paper is as follows. In Section \ref{calcolo}, we analytically work out the PL orbital precessions due to $\xi$. Section \ref{guardami} is devoted to the confrontation with the observations. In Section \ref{precessioni}, we use the secular precessions of the perihelia of some planets of the Solar System. The tests performed with the Sun's angular momentum are critically analyzed in Section \ref{trottola}.
In Section \ref{concludi}, we summarize our findings and draw our conclusions.
\section{Calculation of the orbital effects}\lb{calcolo}
The PL reduced Lagrangian for a localized two-body system of total mass $M$ and extension $r$ is
\eqi\mathcal{L}_{\xi} = \rp{\xi U_G GM}{2c^2 r^3}\ton{\bds r\bds\cdot\bds{{\hat{k}}}}^2.\lb{Lpert}\eqf In it,  $\xi$ is the weak-field Whitehead parameter, $U_G= \kappa \Theta_0^2$ is the Galactic potential at the system's position, assumed proportional to the square of the Galactic rotation velocity $\Theta_0$ at the system's position, $G$ is the Newtonian constant of gravitation, $c$ is the speed of light in vacuum,  $\bds r$ is the two-body relative position vector, and $\bds{\hat{k}}$ is the unit vector pointing from the system to the Galactic Center (GC); in Celestial coordinates,\footnote{The ecliptic coordinates of the GC are  $\lambda_{\rm GC} = 183.15^{\circ}$, $\beta_{\rm GC} = -5.61^{\circ}$ \citep{2004ApJ...616..872R}.} it is
\begin{align}
\kx \lb{kx}& = -0.994,\\ \nonumber \\
\ky & = -0.011,\\ \nonumber \\
\kz \lb{kz}& = -0.111.
\end{align}
The orbital effects induced by \rfr{Lpert}  can be perturbatively computed with the aid of the standard Lagrange  planetary equations  for the variation of the osculating Keplerian orbital elements which are the semimajor axis $a$, the eccentricity $e$, the inclination $I$, the node $\Om$ and the longitude of pericenter $\varpi$.
In general, for a given perturbing Lagrangian $\mathcal{L}_{\rm pert}$, the corresponding perturbing Hamiltonian is \citep{2004A&A...415.1187E,2005CeMDA..91...75E}
\eqi  \mathcal{H}_{\rm pert} = - \mathcal{L}_{\rm pert} -\rp{1}{2}\ton{\derp{{\mathcal{L}}_{\rm pert}}{\bds v}}^2,\eqf where $\bds v$ is the  velocity of the relative orbital motion.
Since \rfr{Lpert} is independent  of $\bds v$, the perturbing Hamiltonian, to be used as disturbing function in the Lagrange equations, is simply
\eqi  \mathcal{H}_{\rm pert} = - \mathcal{L}_{\xi}.\lb{Upert}\eqf
\textcolor{black}{In order to obtain the long term rates of change of the Keplerian orbital elements, $\mathcal{H}_{\xi}$ must be averaged over one orbital period $P_{\rm b} = 2\pi/\nk=2\pi\sqrt{a^3/GM}$. It can be done by evaluating \rfr{Upert} onto the unperturbed Keplerian ellipse, assumed as  reference trajectory, by means of \citep{befa,2011rcms.book.....K}
\begin{align}
x & = r\ton{\cO\cu - \cI\sO\su}, \\ \nonumber \\
y & = r\ton{\sO\cu + \cI\cO\su}, \\ \nonumber \\
z & = r\sI\su, \\ \nonumber \\
r \lb{erre} &= a\ton{1 - e\cos E}, \\ \nonumber \\
dt \lb{dt} & = \ton{\rp{1-e\cos E}{\nk}}dE, \\ \nonumber \\
\sin f & = \rp{\sqrt{1-e^2}\sin E}{1 - e\cos E}, \\ \nonumber \\
\cos f \lb{cosf} & = \rp{\cos E - e}{1- e \cos E}.
\end{align}
In \rfr{erre}-\rfr{cosf}, $E$ is the eccentric anomaly, whose use turns out to be computationally more convenient, $f$ is the\footnote{\textcolor{black}{Sometimes, it is denoted by $\nu$.}} true anomaly, and $u\doteq \omega + f$ is the argument of latitude.
As a result \textcolor{black}{of the averaging process of \rfr{Upert}}, one gets
\begin{align}
\ang{{\mathcal{H}}_{\xi}} \nonumber \lb{Uav} & = -\rp{\xi U_{\rm G} \nk^2 a^2}{c^2 e^2}\grf{-\kx^2\sqrt{1-e^2} \cos ^2 I  \sin ^2\omega \sin ^2\Om + \right. \\ \nonumber \\
\nonumber & + \left. \kx^2\sqrt{1-e^2} \cos  I  \sin  2\omega  \sin  2\Om  - \right. \\ \nonumber \\
\nonumber & - \left. 2 \kx\ky\sqrt{1-e^2} \cos  I  \cos  2\Om  \sin  2\omega +\right.\\\nonumber \\
\nonumber & + \left. \kx\kz\sqrt{1-e^2} \sin  2I  \sin ^2\omega \sin \Om + \right. \\ \nonumber \\
\nonumber & +\left. \kx\ky\sqrt{1-e^2} \cos ^2 I  \sin ^2\omega \sin  2\Om   + \right.\\\nonumber \\
\nonumber & + \left.\kx\ky\sqrt{1-e^2}  \sin ^2\omega \sin  2\Om + \right. \\ \nonumber \\
\nonumber &+ \left.\left[\kx^2\left(e^2+\sqrt{1-e^2}-1\right) -\right.\right.\\ \nonumber \\
\nonumber &-\left.\left. \ky^2\left(\sqrt{1-e^2}-1\right)  \cos ^2 I \right] \cos ^2\Om  \sin ^2\omega - \right. \\ \nonumber \\
\nonumber &-\left. \kz^2\sqrt{1-e^2} \sin ^2 I  \sin ^2\omega +\ky^2\sqrt{1-e^2}  \sin ^2\omega \sin ^2\Om -\right. \\ \nonumber \\
\nonumber &- \left. \ky \cos \Om  \sin ^2\omega\left[\kz\left(\sqrt{1-e^2}-1\right)  \sin  2I +\right.\right. \\ \nonumber \\
\nonumber & + \left.\left. \kx \left(-2 e^2+\cos  2I +3\right) \sin \Om \right]+\sin ^2\omega \left(\kz^2\sin ^2 I +\right.\right. \\ \nonumber \\
\nonumber & +\left.\left. \sin \Om  \left(\left(\ky^2\left(e^2 - 1\right) +\kx^2 \cos ^2 I \right) \sin \Om -\kx \kz \sin  2I \right)\right)+\right.\\ \nonumber \\
\nonumber &+ \left. \ky \cos  I  \sin  2\omega  \left(2 \kx \cos  2\Om +\ky \sin  2\Om \right)+\right.\\ \nonumber \\
\nonumber & + \left.\cos ^2\omega \left(-2 \kx \kz\cos  I  \sin  I  \sin \Om  e^2+ \right.\right. \\ \nonumber \\
\nonumber &+\left.\left.\left(\ky^2\left(e^2+\sqrt{1-e^2}-1\right)  \cos ^2 I-\right.\right.\right.\\ \nonumber \\
\nonumber &- \left.\left.\left. \kx^2\left(\sqrt{1-e^2}-1\right) \right)\cos ^2\Om + \right.\right. \\ \nonumber \\
\nonumber & +\left.\left. \kz^2\left(e^2+\sqrt{1-e^2}-1\right)  \sin ^2 I -\right.\right.\\ \nonumber \\
\nonumber &-\left.\left. \ky^2\sqrt{1-e^2}  \sin ^2\Om + \kx^2\sqrt{1-e^2}  \cos ^2 I  \sin^2\Om +\right.\right. \\ \nonumber \\
\nonumber &+\left.\left. \left(\ky^2+\left(e^2-1\right) \kx^2 \cos ^2 I \right) \sin ^2\Om +\right.\right.\\ \nonumber \\
\nonumber &+\left.\left. \kx \kz\ton{1-\sqrt{1-e^2}}  \sin  2I  \sin \Om  +\right.\right.\\ \nonumber \\
\nonumber &+ \left.\left. \ky\kz \cos \Om  \left(\left(e^2+\sqrt{1-e^2}-1\right) \sin  2I -\right.\right.\right.\\ \nonumber \\
\nonumber &-\left.\left.\left. \kx \left(2 e^2\cos ^2 I +\sqrt{1-e^2} (\cos  2I +3)\right) \sin \Om \right)+\right.\right. \\ \nonumber \\
\nonumber &+\left.\left. \kx \ky \left(\cos ^2 I +1\right) \sin  2\Om \right)+\right.\\ \nonumber \\
\nonumber & +\left. 2 \sin 2\omega \left(\frac{1}{2} \cos  I  \left(\left(\kx^2\left(e^2-2\right) - \right.\right.\right.\right. \\ \nonumber \\
\nonumber &-\left.\left.\left.\left.\ky^2\left(e^2+2 \sqrt{1-e^2}\right) \right) \sin  2\Om - 2e^2 \kx \ky \cos  2\Om \right) -\right.\right. \\ \nonumber \\
& -\left.\left. \kz\left(e^2 + 2 \sqrt{1-e^2}-2\right)  \sin  I  (\kx \cos \Om +\ky \sin \Om )\right) }.
\end{align}
The  long-term orbital rates of change  can, thus, be straightforwardly obtained from \rfr{Uav} by taking simple partial derivatives of it with respect to the orbital elements. Indeed, the planetary Lagrange equations for their averaged orbital rates  are \citep{befa,2011rcms.book.....K}
\begin{align}
\ang{\dot a} \lb{dota} &= -\rp{2}{n_{\rm b}a}\derp{\ang{{\mathcal{H}}_{\rm pert}}}{\mathcal{M}_0}, \\ \nonumber \\
\ang{\dot e} &= \rp{1}{\nk a^2}\ton{\rp{1 - \ee}{e}}\qua{\rp{1}{\sqrt{1-\ee}} \derp{ \ang{ {\mathcal{H}}_{\rm pert} } }{\omega} -
\derp{ \ang{ {\mathcal{H}}_{\rm pert} } }{{\mathcal{M}_0}} }, \\ \nonumber \\
\ang{\dot I} &= \rp{1}{n_{\rm b}a^2\sqrt{1-\ee}\sI}\qua{\derp{\ang{{\mathcal{H}}_{\rm pert}}} {\mathit{\Omega}} - \cI\derp{\ang{{\mathcal{H}}_{\rm pert}}} \omega}, \\ \nonumber \\
\ang{\dot\Om} &= -\rp{1}{n_{\rm b}a^2\sqrt{1-\ee}\sI}\derp{\ang{{\mathcal{H}}_{\rm pert}}} I, \\ \nonumber \\
\ang{\dot\varpi} \lb{dotpi}&= -\rp{1}{n_{\rm b}a^2}\qua{\ton{\rp{\sqrt{1-\ee}}{e}}\derp{\ang{{\mathcal{H}}_{\rm pert}}} e + \rp{\tan\ton{\rp{I}{2}}}{\sqrt{1-\ee}}\derp{\ang{{\mathcal{H}}_{\rm pert}}} I },
\end{align}
where $\mathcal{M}_0$ is the mean anomaly at the epoch. Using  \rfr{Uav} in \rfr{dota}-\rfr{dotpi}  yields
 }
\begin{align}
\ang{\dot a}_{\xi} \lb{dadt} & = 0, \\ \nonumber \\
\ang{\dot e}_{\xi} \lb{dedt} \nonumber & = -\rp{\xi U_G n_{\rm b}\sqrt{1-\ee}\ton{- 2 + \ee + 2\sqrt{1-\ee}}}{4 c^2  e^3  }\textcolor{black}{\times} \\ \nonumber \\
\nonumber &\textcolor{black}{\times}  \grf{ -8 \kz \sI\coo\ton{\kx\cO + \ky\sO} +\right. \\ \nonumber \\
\nonumber & + \left. 4\cI\coo\qua{-2\kx \ky\cOO + \ton{\kx^2 - \ky^2}\sOO} +\right.\\ \nonumber \\
\nonumber & + \left.  \soo\qua{ \ton{\kx^2 - \ky^2}\ton{3 + \cII}\cOO + \right.\right. \\ \nonumber \\
\nonumber & + \left.\left. 2\sin^2 I\ton{\kx^2 + \ky^2 - 2 \kz^2} - \right.\right.\\ \nonumber \\
\nonumber & -\left.\left. 4\kz\sII\ton{\ky\cO - \kx\sO } + \right.\right. \\ \nonumber \\
          & + \left.\left. 2 \kx \ky \ton{3 + \cII}\sOO}}, \\ \nonumber \\
 \ang{\dot I}_{\xi} \lb{dIdt}\nonumber & = -\rp{\xi U_G n_{\rm b}}{ c^2  \ee \sqr }\qua{\kz\cI - \sI\wmn}\textcolor{black}{\times} \\ \nonumber \\
\nonumber &\textcolor{black}{\times}\grf{- 2\sqr\kx\cO\sin^2\omega + \right. \\ \nonumber  \\
\nonumber & + \left. 2\sqr\ky\cI\cO\soo - \right.\\ \nonumber \\
\nonumber & - \left. 2\sqr\ky\sO\sin^2\omega + \right. \\ \nonumber \\
\nonumber & + \left. 2\sqr\cos^2\omega\wpl - \right. \\ \nonumber \\
\nonumber & - \left. 2\coo\wpl - \right. \\ \nonumber \\
\nonumber & - \left. 2\ee\sin^2\omega\wpl + \right. \\ \nonumber \\
\nonumber & + \left. 2\sqr\soo\qua{\kz\sI -\kx\cI\sO} + \right. \\ \nonumber \\
\nonumber & + \left. \ton{-2 + \ee} \soo\qua{\kz\sI + \right.\right.\\ \nonumber \\
          & + \left.\left. \cI\wmn}}, \\ \nonumber \\
\ang{\dot\Om}_{\xi} \lb{dOdt} \nonumber & =  \rp{\xi U_G n_{\rm b}\csc I}{ c^2  \ee\sqr  }\qua{\kz\cI - \sI\wmn}\textcolor{black}{\times}\\ \nonumber \\
\nonumber &\textcolor{black}{\times} \grf{\kz\sI\qua{\ee + \ple\coo} + \right. \\ \nonumber \\
\nonumber & + \left.\cI\qua{\ee + \ple\coo}\textcolor{black}{\times}\right.\\ \nonumber \\
\nonumber &\textcolor{black}{\times}\left. \wmn - \right. \\ \nonumber \\
          & - \left. \ple\soo\wpl}, \\ \nonumber \\
\ang{\dot\varpi}_{\xi} \lb{dvarpidt} \nonumber & = -\rp{\xi U_G n_{\rm b}}{4 c^2  e^4  }\textcolor{black}{\times}\\ \nonumber \\
\nonumber &\textcolor{black}{\times} \grf{\ple\textcolor{black}{\times}\right.\\ \nonumber  \\
\nonumber &\textcolor{black}{\times}  \left.\qua{\wq\cp\cOO\coo + \right.\right. \\ \nonumber \\
\nonumber & + \left.\left. 2\wl\sin^2 I\coo - \right.\right.\\ \nonumber \\
\nonumber & - \left.\left. 4\kz\sII\wmn\coo + \right.\right.\\ \nonumber \\
\nonumber & + \left.\left. 2\kx\ky\cp\sOO\coo + \right.\right.\\ \nonumber \\
\nonumber & + \left.\left. 8\kz\sI\wpl\soo + \right.\right.\\ \nonumber \\
\nonumber & + \left.\left. 8\kx\ky\cI\cOO\soo - \right.\right.\\ \nonumber \\
\nonumber & - \left.\left. 4\wq\cI\sOO\soo} - \right. \\ \nonumber \\
\nonumber & - \left. \rp{4\ee\tan\ton{\rp{I}{2}}}{\sqr}\qua{\kz\cI - \sI\wmn}\textcolor{black}{\times}\right.\\ \nonumber \\
\nonumber &\textcolor{black}{\times} \left.\qua{\kz\ton{\ee + \ple\coo}\sI + \right.\right. \\ \nonumber \\
\nonumber & + \left.\left. \cI\ton{\ee + \ple\coo}\textcolor{black}{\times}\right.\right.\\ \nonumber \\
\nonumber &\textcolor{black}{\times}\left.\left. \wmn - \right.\right.\\ \nonumber \\
          & - \left.\left. \ple\wpl\soo  }}.
\end{align}
In calculating \rfr{dadt}-\rfr{dvarpidt}, we did not make any a-priori assumptions about both the orbital geometry and the spatial orientation of $\bds{\hat{k}}$, which was kept fixed in the integration of \rfr{Upert} over $P_{\rm b}$.  From \rfr{dadt}-\rfr{dvarpidt}, it can be noted that the precessions are proportional to the orbital frequency $\nk$; the tighter the binary system is, the stronger the PL effects are.

In the limit of small eccentricity, \rfr{dadt}-\rfr{dvarpidt} reduce to
\begin{align}
\ang{\dot a}_{\xi} \lb{dadt2} & = 0, \\ \nonumber \\
\ang{\dot e}_{\xi} \lb{dedt2} \nonumber & \sim \rp{\xi U_G n_{\rm b}e}{16 c^2    }\textcolor{black}{\times} \\ \nonumber \\
\nonumber &\textcolor{black}{\times}  \grf{ -8 \kz \sI\coo\ton{\kx\cO + \ky\sO} +\right. \\ \nonumber \\
\nonumber & + \left. 4\cI\coo\qua{-2\kx \ky\cOO + \ton{\kx^2 - \ky^2}\sOO} +\right.\\ \nonumber \\
\nonumber & + \left.  \soo\qua{ \ton{\kx^2 - \ky^2}\ton{3 + \cII}\cOO + \right.\right. \\ \nonumber \\
\nonumber & + \left.\left. 2\sin^2 I\ton{\kx^2 + \ky^2 - 2 \kz^2} - \right.\right.\\ \nonumber \\
\nonumber & -\left.\left. 4\kz\sII\ton{\ky\cO - \kx\sO } + \right.\right. \\ \nonumber \\
          & + \left.\left. 2 \kx \ky \ton{3 + \cII}\sOO}} + \mathcal{O}\ton{e^5}, \\ \nonumber \\
\ang{\dot I}_{\xi} & \sim \rp{\xi U_G \nk}{c^2}\ton{\kx\cO + \ky\sO}\textcolor{black}{\times} \\ \nonumber \\
          & \textcolor{black}{\times} \qua{\kz\cI + \sI\ton{\kx\sO - \ky\cO} } + \mathcal{O}\ton{\ee}, \\ \nonumber \\
\ang{\dot\Om}_{\xi} \lb{dOdt2} \nonumber & \sim  \rp{\xi U_G\nk\csc I}{2  c^2}\qua{\kz\sI + \cI \ton{\ky\cO - \kx\sO} }\textcolor{black}{\times}\\ \nonumber \\
          &\textcolor{black}{\times} \qua{\kz\cI + \sI \ton{\kx\sO - \ky\cO} } + \mathcal{O}\ton{\ee}, \\ \nonumber \\
\ang{\dot\varpi}_{\xi} \lb{dvarpidt2} \nonumber & \sim -\rp{\xi U_G\nk}{4 c^2}\grf{ -\rp{1}{4}\coo\qua{\ton{\kx^2-\ky^2}\ton{3 + \cII}\cOO + \right.\right.\\ \nonumber \\
\nonumber & + \left.\left. 2\sin^2 I\ton{\kx^2 + \ky^2 - 2\kz^2} - \right.\right.\\ \nonumber \\
\nonumber & - \left.\left. 4\kz\sII \ton{\ky\cO -\kx\sO} + \right.\right.\\ \nonumber \\
\nonumber & + \left.\left. 2\kx\ky \ton{3 + \cII}\sOO} + \right.\\ \nonumber \\
\nonumber & +\left. 2\qua{-\kz\sI +\cI\ton{\kx\sO - \ky\cO} }\textcolor{black}{\times} \right.\\ \nonumber \\
\nonumber & \textcolor{black}{\times} \left. \qua{ \soo\ton{\kx\cO + \ky\sO} + 2\kz\cI\tan\ton{\rp{I}{2}} +\right.\right.\\ \nonumber \\
          & +\left.\left. 2\sI\ton{\kx\sO - \ky\cO}\tan\ton{\rp{I}{2}}   }  } + \mathcal{O}\ton{\ee}.
\end{align}
From \rfr{dadt2}-\rfr{dvarpidt2}, it can be noticed that, while for the eccentricity the first non-vanishing term is of order $\mathcal{O}\ton{e}$, it is of order zero in $e$ for the other elements.
\section{Confrontation with the observations}\lb{guardami}
\subsection{The perihelion precessions}\lb{precessioni}
The exact expressions of Section \ref{calcolo} can be applied, in principle, to quite different scenarios involving arbitrary orbital configurations characterized by, e.g., high eccentricities such as specific artificial satellites, major bodies of the Solar System, exoplanets, compact binaries, etc.

Let us, now, focus on the Solar System's planets. Latest studies in the field of planetary ephemerides \citep{2011CeMDA.111..363F,2013MNRAS.432.3431P} point towards the production of supplementary secular precessions of more than one orbital element for an increasing number of planets. At present, we have at our disposal the extra-precessions $\Delta\dot\Om, \Delta\dot\varpi$  for Mercury, Venus, Earth, Mars, Jupiter, Saturn \citep{2011CeMDA.111..363F,2013MNRAS.432.3431P}; see Table \ref{tavola}.
\begin{table*}
\caption{Planetary periehlion and node extra-precessions $\Delta\dot\varpi,\Delta\dot\Om$, in milliarcseconds per century (mas cty$^{-1}$), estimated by \citet{2011CeMDA.111..363F} with the INPOP10a ephemerides and by \citet{2013MNRAS.432.3431P} with the EPM2011 ephemerides. In both cases, the general relativistic gravitomagnetic field of the Sun was not modelled, contrary to  all the other known dynamical effects of classical and relativistic origin.  The resulting Lense-Thirring precessions are negligible for all the planets, apart from Mercury \citep{2012SoPh..281..815I}. The values of all the PPN parameters were kept fixed to their general relativistic values; thus, no PL effects were modelled at all. The supplementary perihelion precessions of Venus and Jupiter determined by \citet{2013MNRAS.432.3431P} are non-zero at the $1.6\sigma$ and $2\sigma$ level, respectively.
}\label{tavola}
\centering
\bigskip
\begin{tabular}{llll}
\hline\noalign{\smallskip}
Planet & $\Delta\dot\varpi$ \citep{2013MNRAS.432.3431P}& $\Delta\dot\varpi$ \citep{2011CeMDA.111..363F} & $\Delta\dot\Om$ \citep{2011CeMDA.111..363F} \\
\noalign{\smallskip}\hline\noalign{\smallskip}
Mercury & $-2\pm 3$ &  $0.4 \pm 0.6$ & $1.4 \pm 1.8$ \\
Venus & $2.6\pm 1.6$ &  $0.2\pm 1.5$ & $0.2\pm 1.5$ \\
Earth & $0.19 \pm 0.19$ & $-0.2\pm 0.9$ & $0.0\pm 0.9$ \\
Mars & $ -0.020\pm 0.037$ & $-0.04 \pm 0.15$ & $-0.05\pm 0.13$ \\
Jupiter & $ 58.7\pm 28.3$ & $-41\pm 42$ & $-40\pm 42$ \\
Saturn & $-0.32\pm 0.47$ & $0.15\pm 0.65$ & $-0.1\pm 0.4$ \\
\noalign{\smallskip}\hline\noalign{\smallskip}
\end{tabular}
\end{table*}
Such extra-rates are accurate to the $\sim 1-0.04$ milliarcseconds per century (mas cty$^{-1}$) level, apart from Jupiter for which a $\sim 30-40$ mas cty$^{-1}$ accuracy level has been reached so far. They are to be intended as corrections to the standard planetary precessions since they were determined in fitting increasingly accurate dynamical models  to data records for the major bodies of the Solar System covering about one century. The theoretical expressions used in such big  data reductions account for an almost complete set of dynamical effects caused by the currently accepted laws of gravitation. In particular, GR is fully modeled, with the exception of the gravitomagnetic Lense-Thirring (LT) effect \citep{LT18} induced by the Sun's angular momentum $\bds S$. Thus, in principle, $\Delta\dot\Om$ and $\Delta\dot\varpi$ account for any mismodelled/unmodelled dynamical effect having an impact on the planetary orbital motions within the current accuracy. At present, some of them are  statistically compatible with zero \citep{2011CeMDA.111..363F,2013MNRAS.432.3431P}. As such, they can can be suitably used in non-detection tests to check the hypothesis  $\xi = 0$ by inferring \textcolor{black}{preliminary} upper bounds on it from a direct comparison to our theoretical predictions in Section \ref{calcolo}. For a general discussion of such an approach, widely followed in the literature to constrain a variety of non-standard effects with existing data, see Section $4.2$ of \citet{2012arXiv1210.3026I} and references therein. Here, we briefly point out that, strictly speaking, the present approach allows to test alternative theories differing from  GR just for $\xi$, being all the other PPN parameters set to their GR values. If, in future studies, the astronomers will explicitly model the PL dynamical effects, then, $\xi$ could be simultaneously estimated along with a selection of other PPN parameters. \textcolor{black}{Such an approach would allow to  put effective constraints on $\xi$.}
In calculating \rfr{dvarpidt}, it is necessary to evaluate $U_G$. To this aim, we will recur to\textcolor{black}{, e.g.,} the recent model by \citet{Cinesi013}. It consists of the following components for the Galactic potential
\begin{align}
U_{\rm bulge} & = -\rp{GM_{\rm b}R}{\ton{R + R_{\rm b}}^2}, \\ \nonumber \\
U_{\rm disk} & = -4\pi G\Sigma_{\rm d} R_{\rm d} \eta^2\qua{I_0\ton{\eta} K_0\ton{\eta}- I_1\ton{\eta}K_1\ton{\eta}} \\ \nonumber \\
U_{\rm halo} & = -\sqrt{4\pi G\rho R_{\rm h}^2}\qua{1 - \rp{R_{\rm h}}{R}\arctan\ton{\rp{R}{R_{\rm h}} } },
\end{align}
where $R$ is the distance from the GC,
$M_{\rm b}$ is the mass of the bulge, $R_{\rm b}$ is the half-mass scale radius of the bulge, $\Sigma_{\rm d}$ is the central surface mass density of the disk, $R_{\rm d}$ is the scale radius of the disk,
\eqi \eta\doteq \rp{R}{2R_{\rm d}},\eqf $I_j\ton{\eta}, K_j\ton{\eta}$ are the modified Bessel function of the first and second kind, respectively, $\rho$ is the central volume mass density of the halo, $R_{\rm h}$ is the scale radius of the halo. Their values are listed in Table \ref{tavola2}.
\begin{table*}
\caption{Parameters of the model adopted  for the Galactic potential, from Table 3 in \citet{Cinesi013}. $M_{\rm b}$ is the mass of the bulge, $R_{\rm b}$ is the half-mass scale radius of the bulge, $\Sigma_{\rm d}$ is the central surface mass density of the disk, $R_{\rm d}$ is the scale radius of the disk,
 $\rho$ is the central volume mass density of the halo, $R_{\rm h}$ is the scale radius of the halo.
}\label{tavola2}
\centering
\bigskip
\begin{tabular}{lll}
\hline\noalign{\smallskip}
Parameter & Value & Units  \\
\noalign{\smallskip}\hline\noalign{\smallskip}
$M_{\rm b}$ & $1.5\times 10^{10}$ & M$_{\odot}$ \\
$R_{\rm b}$ & $0.3$ & kpc\\
$\Sigma_{\rm d}$ & $1.2\times 10^3$ & M$_{\odot}$ pc$^{-2}$ \\
$R_{\rm d}$ & $4.0$ & kpc \\
$\rho$ & $0.01$ & M$_{\odot}$ pc$^{-3}$ \\
$R_{\rm h}$ & $5.0$ & kpc \\
\noalign{\smallskip}\hline\noalign{\smallskip}
\end{tabular}
\end{table*}
By adopting \citep{2009ApJ...700..137R, Cinesi013}
\begin{align}
\Theta_0 & = 254\ {\rm km\ s^{-1}},\\ \nonumber \\
R_{\odot} & = 8.4\ {\rm kpc},
\end{align} it turns out
\eqi \rp{U_G}{c^2} = 6.4\times 10^{-7},\lb{galpot}\eqf
implying
\eqi\kappa = 0.89.\lb{kappazzo}\eqf
\textcolor{black}{Many other models for the Galactic potential are available in the literature; see, e.g., \citet{1990ApJ...348..485P, 2010MNRAS.402..934M, 2010MNRAS.408.1788B,2011MNRAS.414.2446M,  2012PASJ...64..136H}. Nonetheless, it turns out that their use does not have a significant impact on our results.
As an example,} \citet{2013arXiv1307.2637S}, by adopting the Galaxy potential model in \citet{1990ApJ...348..485P}, obtained
\eqi \rp{U_G}{c^2} = 5.4\times 10^{-7}.\eqf
From the latest results by \citet{2013MNRAS.432.3431P}, and by using \rfr{kx}-\rfr{kz} and \rfr{galpot}, it turns out that the tightest \textcolor{black}{bound} on $\xi$ comes from the perihelion of Mars;
\eqi |\xi|\leq 5.8\times 10^{-6}.\lb{constraint1}\eqf
The \textcolor{black}{preliminary bound} of \rfr{constraint1}, which is three orders of magnitude better than \textcolor{black}{the constraint} obtained by\footnote{\textcolor{black}{Caution is advised in straightforwardly compare bounds such as the one in \rfr{constraint1} and in \citet{1987ApJ...320..871N} with the constraint by \citet{1976ApJ...208..881W}  which is the outcome of an actual lab-type experiment.}} \citet{1976ApJ...208..881W} from superconducting gravimetry, is at the same level of the \textcolor{black}{bound} inferred by \citet{2013arXiv1307.2637S} from the Sun's spin axis evolution; see Section \ref{trottola} for a critical discussion of this point.
%
%
%
%
%

\textcolor{black}{It is worthwhile noticing that the bound of \rfr{constraint1} is close to the  constraints, of the order of just $10^{-6}$, expected from different versions of the radioscience experiment of the future BepiColombo mission to Mercury \citep{2007PhRvD..75b2001A}. }
As shown by Table \ref{tavola},  \citet{2013MNRAS.432.3431P}, using the EPM2011 ephemerides, obtained non-zero supplementary perihelion precessions  for Venus and Jupiter, although  not at a $\geq 3\sigma$ level. This allows us to preliminarily take the ratio of the perihelia of such planets, thus setting up a test of the hypothesis $\xi \neq 0$, irrespectively of its actual value and of $U_G$ itself as well. From Table \ref{tavola}, it turns out
\begin{align}
\rp{\Delta\dot\varpi_{\rm Ven}}{\Delta\dot\varpi_{\rm Jup}} \lb{ratio1}& = 0.044\pm 0.034.
%
%
\end{align}
From \rfr{dvarpidt},
one has for the ratio of the PL perihelion precessions of Venus and Jupiter
\begin{align}
\rp{\ang{\dot\varpi}_{\xi}^{\rm Ven}}{\ang{\dot\varpi}_{\xi}^{\rm Jup}} \lb{test1} & = 9.148.
%
%
\end{align}
 It is \textcolor{black}{clearly} incompatible with \rfr{ratio1}, independently of $\xi$ (assumed non-zero). Nonetheless, further astronomical investigations are required to confirm or disproof the existence of the non-zero precessions for Venus and Jupiter as  trustable anomalies needing explanation. Caution is in order in view of the fact that, e.g., the anomalous perihelion precession for Saturn, independently reported a few years ago by two teams of astronomers with the INPOP08 and EPM2008 ephemerides at the $1.2-3\sigma$ level\footnote{See Table 4 in \citet{2010IAUS..261..159F}, and Table 5 in \citet{2011CeMDA.111..363F}. See also \citet{2009AJ....137.3615I}, where possible causes were discussed.}, was not confirmed in further studies \citep{2010IAUS..261..170P,2011CeMDA.111..363F,2013MNRAS.432.3431P}.
\subsection{The precession of the Sun's spin axis}\lb{trottola}
The orientation of the  Solar System's invariable plane \citep{2012A&A...543A.133S} is determined by
the values of the celestial coordinates of its north pole which, at the epoch J2000.0, are \citep{2007CeMDA..98..155S}
\begin{align}
\alpha_{\rm inv} & = 273.85^{\circ},\\ \nonumber \\
\delta_{\rm inv} & = 66.99^{\circ}.
\end{align}
Its normal unit vector $\bds{\hat{L}}$, in Celestial coordinates, is
\begin{align}
\hat{L}_x & = \cos\delta_{\rm inv}\cos\alpha_{\rm inv} = 0.03, \\ \nonumber \\
\hat{L}_y & = \cos\delta_{\rm inv}\sin\alpha_{\rm inv} = -0.39, \\ \nonumber \\
\hat{L}_z & = \sin\delta_{\rm inv} = 0.92.
\end{align}
The north pole of rotation of the Sun at the epoch J2000.0 is  characterized by
\citep{2007CeMDA..98..155S}
\begin{align}
\alpha_{\odot} & = 286.13^{\circ},\\ \nonumber \\
\delta_{\odot} & = 63.87^{\circ},
\end{align}
so that the Sun's spin axis $\bds{\hat{S}}$, in Celestial coordinates, is
\begin{align}
\hat{S}_x \lb{spinx} & = 0.12, \\ \nonumber \\
\hat{S}_y \lb{spiny} & = -0.43, \\ \nonumber \\
\hat{S}_z \lb{spinz} & = 0.89.
\end{align}
Thus, the angle  between the Sun's spin axis $\bds{\hat{S}}$ and the Solar System's total angular momentum $\bds{L}$ is
\eqi\theta=5.97^{\circ}.\lb{angolo}\eqf

\citet{1987ApJ...320..871N} \textcolor{black}{indirectly obtained} \eqi |\xi|\lesssim 10^{-7}\lb{Nor}\eqf  from \rfr{angolo} and the Sun's spin axis evolution. Similarly, \citet{2013arXiv1307.2637S},  inferred
\eqi |\xi|\lesssim 5\times 10^{-6}.\lb{Cin}\eqf
Basically, the reasonings by \citet{1987ApJ...320..871N} and \citet{2013arXiv1307.2637S} rely upon the fact that  $\theta$  changes with time because of the PL precession of $\bds{\hat{S}}$ around $\bds{\hat{k}}$ at a rate $\Psi_{\xi}$ proportional to $\xi$. Under certain simplifying assumptions,  \citet{1987ApJ...320..871N} obtained
\eqi\sin\ton{\rp{\theta}{2}}=\ton{\rp{\Psi_{\xi}}{\Psi_{\rm class}}}\sin\ton{\rp{\Psi_{\rm class} t}{2}},\lb{torquo}\eqf where $\Psi_{\rm class}\sim 10^{-10}$ yr$^{-1}$  is the precession rate of $\bds{\hat{S}}$ due to the classical torques exerted by the major planets of the Solar System. By requiring that the Sun's equator and the invariable plane were closely aligned at the birth of the Solar System $\sim 5$ Gyr ago, and by assuming the steady action of both the PL and the classical torques throughout the life of the Solar System in such a way that $\Psi_{\rm class} \Delta T/2\ll 1, \Delta T=5\ {\rm Gyr}$, \rfr{torquo} reduces to
\eqi \theta \sim \Psi_{\xi}\Delta T.\lb{limite}\eqf Thus, the PL spin precession would cause a secular rate of change $\dot\theta$ whose magnitude is approximately equal just to $\Psi_{\xi}$. By posing $\Delta T=5\ {\rm Gyr}$, \citet{1987ApJ...320..871N} inferred an upper limit of $\xi$
as little as $|\xi|\lesssim 10^{-7}$ by comparing $\Psi_{\xi}$ to the hypothesized value of the secular rate of change of $\theta$
\eqi\dot\theta=\rp{5.97^{\circ}}{5\ {\rm Gyr}} = 0.4\ {\rm mas\ cty^{-1}}.\lb{pazzo}\eqf

Let us, now, propose some critical considerations about how \rfr{Nor}-\rfr{Cin} were obtained.
\begin{enumerate}
\item
The bounds by \citet{1987ApJ...320..871N} and \citet{2013arXiv1307.2637S}  strongly depend on the assumption that the two angular momenta were aligned just after the formation of the Solar System  $\sim 5$ Gyr ago.   If compared to the  well known Solar System's planetary orbital motions and the dynamical forces characterizing them,  such an assumption is  more speculative and  hardly testable since, at present, it seems that there is no empirical information about the long term history of the Sun's obliquity over the eons. Moreover, in recent years several close-in giant planets exhibiting neat misalignments between their orbital angular momentum and their host star's spin axis have been discovered so far \citep{2010ApJ...719..602S,2010ApJ...718L.145W,2011IAUS..276..230W,2012Natur.491..418B}. It would be, perhaps, premature to infer conclusions valid also for our Solar System  in view of the differences among it and the planetary systems in which such a phenomenon has been discovered so far. About multiplanet systems relatively more similar to our Solar System, some of them exhibit low stellar obliquities \citep{2013ApJ...766..101C,2013ApJ...771...11A}; on the other hand, the $110-$day period planet candidate KOI-368.01, whose host star is only $\sim 0.2-0.5$ Gyr old, may have a strong spin-orbit misalignment \citep{2013arXiv1307.2249Z}.
However, the assumption of a primeval close alignment of the Sun's angular momentum with the total angular momentum of the Solar System should, now, be considered as less plausible than before.
\item
Also the Preferred Frame (PF) PPN parameter $\alpha_2$ causes an analogous precession of the angular momentum of an isolated body \citep{1987ApJ...320..871N}. Thus, by looking solely at the time evolution of the Sun's angular momentum, it is impossible to genuinely separate the bounds on $\alpha_2$ and $\xi$, contrary to the case of the planetary orbital precessions. Indeed, depending on the number of planetary extra-precessions at disposal, it is possible  to linearly combine several rates to separate, by construction, the effects of $\xi,\alpha_2$ and, in principle, of other dynamical effects for which analytical expressions of their orbital precessions are available \citep{2012arXiv1210.3026I}.
\item\lb{eiono}
Both \citet{1987ApJ...320..871N} and \citet{2013arXiv1307.2637S} assumed that $\bds{S}$ undergoes a secular precession around $\bds{\hat{k}}$, while $\bds {L}$ stays constant in such a way that $\theta$ changes in time because of the $\xi$-induced precessions of $\bds{S}$ only. Actually, also $\bds{L}$ does the same as $\bds{S}$, although with a longer times scale. Indeed, the  invariable plane is defined as the plane
perpendicular to the total angular momentum vector of the Solar
System that passes through its barycentre \citep{2012A&A...543A.133S}. In turn, the total angular momentum of a $N$-body system such as the Solar System is defined as\footnote{In fact, \citet{1987ApJ...320..871N} did consider the effect of $\xi$ on the orbital angular momentum of each planet, but, relying upon the present-day magnitudes of $S$ and $L_s$, he neglected it.} \citep{2012A&A...543A.133S}
\eqi\bds {L}\doteq\sum_{s=1}^N{\bds L}_s = \sum_{s=1}^N m_s {\bds r}_s\bds\times{\bds v}_s,\eqf where $m_s$, ${\bds r}_s$, and ${\bds v}_s$ are the mass, barycentric position vector,
and barycentric velocity vector of the $s$th body, respectively. As such, the orbital effects of Section \ref{calcolo} necessarily translate into an overall $\xi$-induced variation of $\bds{L}$ itself. When timescales as long as the lifetime of the Solar System are considered, it may have an impact (See point \ref{poveridd}).
\item\lb{gyro}
Another assumption made by \citet{1987ApJ...320..871N} and \citet{2013arXiv1307.2637S} concerns the Sun's rotation rate. Indeed, it was assumed constant throughout the entire Solar lifetime. Actually, it is not so, especially during the early stages after the Sun's formation. Indeed, the fast rotation of a collapsing protostar is slowed down likely because of the magnetic braking caused by  the interaction of the protostar's magnetic field with the stellar wind carrying away angular momentum  \citep{2000MNRAS.312..387F}. For main-sequence stars, the slowing-down of the rotation rate goes as $\sim \tau^{-1/2}$ \citep{Tass72}, where $\tau$ is the stellar age,  as per the Skumanich law \citep{1972ApJ...171..565S}.  Gyrochronology is an established technique to estimate the age of Sun-like like main sequence stars from their rotational period  \citep{2003ApJ...586..464B}.
\item\lb{itemi}
A less speculative and more controlled way to use the Sun's spin to put constraints on $\xi$ from a putative $\dot\theta$ consists in actually monitoring the time evolution of $\theta$ over the years looking for a possible secular variation of it. From a non-detection of $\dot\theta$, namely from a $\Delta\dot\theta$ statistically compatible with zero, upper bounds on $\xi$ can be inferred in as much the same way as we did with the perihelion precessions in Section \ref{precessioni} and \citet{2013arXiv1307.2637S} themselves did from the non-detection of spin precessions in isolated millisecond pulsars. In fact, such a strategy is severely limited by the accuracy with which such a putative $\dot\theta$ could be measured. While the orientation of the invariable plane is nowadays known at the $\sim 0.1-0.3\ {\rm mas\ cty^{-1}}$ level \citep{2012A&A...543A.133S}, to be compared with \rfr{pazzo}, the situation for the Sun's spin axis is less favorable. Its determination is usually made in terms of the Carrington elements
\citep{Carr1863,AA013}
$i$, which is the angle from the Sun's equator to the ecliptic, and the longitude of the node $\mathrm{\Omega}$ of the Sun's equator with respect to the Vernal equinox $\curlyvee$ along the ecliptic, by means of
\begin{align}
\hat{S}_X \lb{Sx} & = \sin i\sin{\mathrm{\Omega}}, \\ \nonumber \\
\hat{S}_Y \lb{Sy} & = -\sin i\cos{\mathrm{\Omega}}, \\ \nonumber \\
\hat{S}_Z \lb{Sz} & = \cos i.
\end{align}
\citet{2005ApJ...621L.153B} recently measured them
from time-distance helioseismology analysis of Dopplergrams from the Michelson Doppler Imager (MDI) instrument on
board the SOlar and Heliospheric Observatory (SOHO) spacecraft; the data span was $\Delta t = 5$ yr, from May 1996  through
July 2001. The resulting values are\footnote{After rotating \rfr{Sx}-\rfr{Sz}, calculated with \rfr{iCarr}-\rfr{OCarr}, from the ecliptic to the Earth's equator, \rfr{spinx}-\rfr{spinz} are obtained.}
\begin{align}
i \lb{iCarr}& = 7.155^{\circ}\pm 0.002^{\circ}, \\ \nonumber \\
{\mathrm{\Omega}} \lb{OCarr}& = 73.5^{\circ}\pm 1^{\circ}.
\end{align}
The figures of \rfr{iCarr}-\rfr{OCarr} imply an ability to observationally constrain a putative rate of change of $\theta\ton{i,{\mathrm{\Omega}},\alpha_{\rm inv},\delta_{\rm inv}}$ with a necessarily limited accuracy; it can approximately be evaluated as
\eqi
\sigma_{\dot\theta}\sim \rp{\sqrt{\ton{\derp{\theta}{i}}^2\sigma^2_{i} + \ton{\derp{\theta}{{\mathrm{\Omega}}}}^2\sigma^2_{{\mathrm{\Omega}}}}}{\Delta t}= 7.73\times 10^2\ {\rm arcsec\ yr^{-1}}.
\eqf
\item\lb{poveridd}
\citet{2013arXiv1307.2637S} integrated $\theta(t)$ backward in time to the epoch $t_{\rm orig} = -4.6$ Gyr for given values of $\xi$. \citet{2013arXiv1307.2637S} used \rfr{angolo} as initial condition common  to all their integrations. Moreover, they also took into account the slow time dependence of $\bds{\hat{k}}$ due to the orbital motion of the Solar System through the Galaxy during its entire lifetime. \citet{2013arXiv1307.2637S} inferred \rfr{Cin} from the condition $\theta_{\rm orig}\lesssim 10^{\circ}$. Actually, as shown at point \ref{itemi}, $\theta$ has an uncertainty of $\sigma_{\theta}\sim 1^{\circ}$; thus, the initial conditions of the integrations for $\theta(t)$ should be varied taking into account $\sigma_{\theta}$ as well. Another potential source of uncertainty in \rfr{Cin} resides in the fact that a realistic propagation of the Sun's orbital motion through the Galaxy backward in time over $\sim 5$ Gyr is not free from uncertainties \citep{2009AN....330..857I,2010MNRAS.403.1469I}. Indeed, it strongly depends on the details of both the baryonic and non-baryonic matter distribution within the Galaxy, and on the dynamical laws adopted. Moreover, due to such factors, the Sun can be finally  displaced even quite far from its current location.
\item
\citet{2013arXiv1307.2637S}, from the non-detection of any anomalous spin precession for some isolated radio pulsars inferred from an analysis of their extremely stable pulse profiles, obtained
\eqi|\hat{\xi}|< 3.9\times 10^{-9}\lb{boundazzo}\eqf for the strong-field version of the Whitehead parameter. As discussed by \citet{2013arXiv1307.2637S}, caution is in order before straightforwardly comparing the bounds on $\hat{\xi}$ to those on $\xi$ because of possible strong-field effects arising in  alternative gravity theories.
The constraint of \rfr{boundazzo} can be converted into an upper  bound on the spatial anisotropy of the Newtonian gravitational constant $G$ as tight as \citep{2013arXiv1307.2637S}
\eqi\left|\rp{\Delta G}{G}\right|^{\rm anisotropy} < 4\times 10^{-16}.\lb{limitazzo}\eqf According to \citet{2013arXiv1307.2637S}, \rfr{limitazzo} is the most constraining limit on the anisotropy of $G$, being four orders of magnitude better than that achievable with LLR in the foreseeable future. By using the same approach followed here, \citet{2011CQGra..28v5027I} obtained
\eqi \left|\rp{\Delta G}{G}\right|^{\rm anisotropy} \lesssim 10^{-17}.\lb{limitazzomio}\eqf
from the non-detection of certain effects in Solar System's planetary orbital motions.
\end{enumerate}
\section{Summary and conclusions}\lb{concludi}
We analytically worked out the orbital perturbations induced in a binary system by a possible violation of the Local Position Invariance taken into account by the PPN Whitehead parameter $\xi$. It turned out that, apart from the semimajor axis $a$, the eccentricity $e$, the inclination $I$, the longitude of the ascending node $\Om$ and the longitude of pericenter $\varpi$ undergo long-term variations. Our calculation is exact in the sense that we did not recur to
a priori simplifying assumptions on the orbital geometry. Also the orientation of the unit vector $\bds{\hat{k}}$ pointing from the binary system to the Galactic Center was not subjected to any restriction. Thus, our results are quite general, and can be applied, in principle, to a variety of different astronomical and astrophysical scenarios.

Following an approach widely adopted in the literature also for other non-standard dynamical effects, we exploited the non-detection of  anomalous perihelion precessions for some planets of the Solar System to indirectly obtain \textcolor{black}{preliminary} upper bounds on $\xi$. \textcolor{black}{Strictly speaking, they may not be regarded as constraints since they do no come from a least-square parameter estimation in a dedicated fit to observations; they should rather be seen as an indication of acceptable values. However, this holds also for other bounds existing in the literature and discussed here}. From the latest orbital determinations for Mars with the EPM2011 ephemerides, and by using recent results on the Galactic potential at the Sun's location, we were able to infer $|\xi|\lesssim 5.8\times 10^{-6}$\textcolor{black}{, which is close to the expected constraints of the order of $10^{-6}$ from the BepiColombo mission to Mercury}. With our
 approach, the hypothesis $\xi = 0$ is tested, and \textcolor{black}{preliminary bounds} on it are posed. A complementary approach which could be followed consists in explicitly modeling the Preferred Localtion effects in the software of the planetary ephemerides and including $\xi$ in the parameters to be estimated in new, dedicated fits to the same planetary data records. The availability of marginally significant non-zero anomalous perihelion precessions for Venus ($1.6\sigma$ level) and Jupiter ($2\sigma$ level) allowed us to put on the test the hypothesis $\xi \neq 0$ by taking the ratio of their precessions. The discrepancy between the theoretical and the observational ratios turned out to be \textcolor{black}{quite large}. However, caution is in order since further analyses are needed to confirm or disproof the real existence of such  potentially interesting astrometric anomalies in the Solar System.

Finally, we critically discussed the $\sim 10^{-6}-10^{-7}$ \textcolor{black}{bounds} from the close alignment of the angular momenta of the Sun and of the Solar System existing in the literature. From a purely phenomenological point of view, $|\xi|\sim 10^{-7}$ would imply an ability to monitor  a secular rate of change of the angle $\theta$ between both the angular momenta with an accuracy of $\sim 0.4$ mas yr$^{-1}$, which is far too small. Indeed, the location of the Sun's spin axis is currently known to the $\sim 1^{\circ}$ level due to the uncertainty in one of the two Carrington elements determining its orientation in space. Such an uncertainty, along with the one coming from the Galactic potential affecting the orbital motion of the Sun throughout the Galaxy over the eons and the fact that main sequence stars rotate faster at the early stages of their life, should be accounted for when the evolution of $\theta$ backward in time is considered to constrain $\xi$. From a broader point of view, at present there are no observations able to provide us with insights concerning the evolution of $\theta$ over the past eons. Moreover, the assumption that $\theta$ at the origin of the Solar System was close to its current value might be now regarded as less likely because of the recent discovery of several exoplanets exhibiting strong spin-orbit misalignment. All in all, the hypotheses and the assumptions on which the spin-type \textcolor{black}{bounds} $|\xi|\sim 10^{-6}-10^{-7}$ rely upon are less established than the determination of the Solar System planetary orbital precessions.
\bibliography{PFEbib,ephemeridesbib}{}

\end{document}